\documentclass[journal=jpclcd,manuscript=letter]{achemso}
\pdfoutput=1
\usepackage{color}
\usepackage{soul}
\sethlcolor{yellow}
\usepackage{tabularx}
\usepackage{graphicx}
\usepackage{bm}
\usepackage{float}
\usepackage[version=3]{mhchem}

\newcommand{\bv}[1]{\mathbf{#1}}

\newcommand{\COMMENT}[1]{}
\newcommand{\textapprox}{{\raise.17ex\hbox{$\scriptstyle\mathtt{\sim}$}}}

\author{Fan Zheng, Liang Z. Tan, Shi Liu, and
 Andrew M. Rappe} 
\affiliation{The Makineni Theoretical Laboratories,
 Department of Chemistry, University of Pennsylvania, Philadelphia,
 Pennsylvania  19104-6323, USA}
\email{rappe@upenn.sas.edu}

\title{Rashba Spin-Orbit Coupling Enhanced Carrier Lifetime in Organometal
 Halide Perovskites} 

\begin{document}

 \begin{abstract} 
   Organometal halide perovskites are promising
   solar-cell materials for next-generation photovoltaic applications.
   The long carrier lifetime and diffusion length of these materials
   make them very attractive for use in light absorbers and carrier transporters.
   While these aspects of organometal halide perovskites have attracted the most attention, the consequences of the Rashba effect, driven by strong spin-orbit coupling, on  the photovoltaic properties of these materials are largely unexplored. 
   In this work, taking the electronic structure of methylammonium lead iodide as an example, we propose an intrinsic mechanism for enhanced carrier lifetime in 3D Rashba materials.  Based on first-principles calculations and a Rashba spin-orbit model, we demonstrate
   that the recombination rate is reduced due to the spin-forbidden
   transition. These results are important for understanding the fundamental
   physics of organometal halide perovskites and for optimizing and designing
   the materials with better performance. The proposed mechanism including spin degrees of freedom offers a new paradigm  of using 3D Rashba materials for photovoltaic applications. 
 \end{abstract}

The organometal halide perovskites (OMHPs) have attracted significant
attention due to the rapid increase in their photovoltaic power conversion
efficiency. In the past 2 years, the reported efficiency of OMHP-based
solar cells has almost doubled from 9.7\%\cite{Kim12p591} to over
20\%~\cite{Zhou14p542,Jeon14p7837,KRICT}, making OHMPs very promising for
low-cost and high-efficiency photovoltaics. Methylammonium lead iodide,
CH$_{3}$NH$_{3}$PbI$_{3}$ (MAPbI$_3$), and other closely-related hybrid
perovskites such as Cl-doped and Br-doped MAPbI$_3$
(MAPbI$_{3-x}$Cl$_{x}$ and MAPbI$_{3-x}$Br$_{x}$),
(NH$_{2}$)$_2$CHPbI$_{3}$ (formamidinium lead iodide, FAPbI$_3$), and
Sn-doped MAPbI$_3$ (MAPb$_{x}$Sn$_{1-x}$I$_{3}$), all display band gaps
(1.1 to 2.1 eV) in the visible light region,  favorable for photovoltaic
applications~\cite{Papavassiliou95p1713,Noel14px,Umari14p4467,Chiarella08p045129,
Ogomi14p1004,Eperon14p982,Stoumpos13p9019,Mosconi14p16137,Filip14p245145}. 
The class of materials also possesses strong light absorption, fast
charge generation and high carrier mobility
\cite{Ponseca14p5189,Edri14p3461}. In particular, exceptionally long
carrier lifetime and diffusion length have been observed in MAPbI$_3$
and MAPbI$_{3-x}$Cl$_{x}$, making them better solar cell candidates than
other semiconductors with similar band gaps and absorption
coefficients~\cite{Xing13p344,Stranks13p341,Wehrenfennig14p1584}.

Intense research has been directed toward understanding and further enhancing the
long carrier lifetime and diffusion length in OMHPs. Previous studies
reported a relatively low defect concentration in
MAPbI$_3$\cite{Kim14p1312,Yin14p063903,Du14p9091,Yamada15p482,
Buin14p6281}, which reduces the scattering centers for nonradiative charge carrier recombination.
Recently, it has been suggested that the spatial carrier segregation caused by
disorder-induced localization\cite{Ma15p248} or domains acting as internal {\em p}-{\em n}
junctions\cite{Frost14p081506,Frost14p2584,Liu15p693} may reduce the recombination rate.
Though many of the OMHPs are 3D Rashba materials driven by strong spin-orbit coupling (SOC) and bulk ferroelectricity\cite{Kim14p6900,Even13p2999,Stroppa14p5900,Amat14p3608}, the effects of spin and orbital degrees of freedom on photovoltaic applications are largely unexplored beyond band gap engineering~\cite{Even13p2999}. 
In this work, we focus on an intrinsic mechanism for the enhancement of long carrier lifetime due to the Rashba splitting.
Using first-principles calculations and effective models,
we find that the Rashba splitting arising from SOC under inversion symmetry breaking can result in spin-allowed and spin-forbidden recombination channels. The spin-forbidden recombination path has a
significantly slower transition rate due to the mismatch of spin and momentum. The spin-allowed recombination path, though kinetically favorable, can be suppressed under appropriate spin texture due to the low population of free carriers. 
Taking the electronic structures of MAPbI$_3$ under various distortions as examples, we show that the proposed mechanism is possible under room temperature, and is potentially responsible for the long carrier lifetime in OMHPs. 
This spin-dependent recombination mechanism highlights the possibility of using 3D Rashba materials for efficient photovoltaic applications.

\begin{figure} \includegraphics[width=1.0\columnwidth]{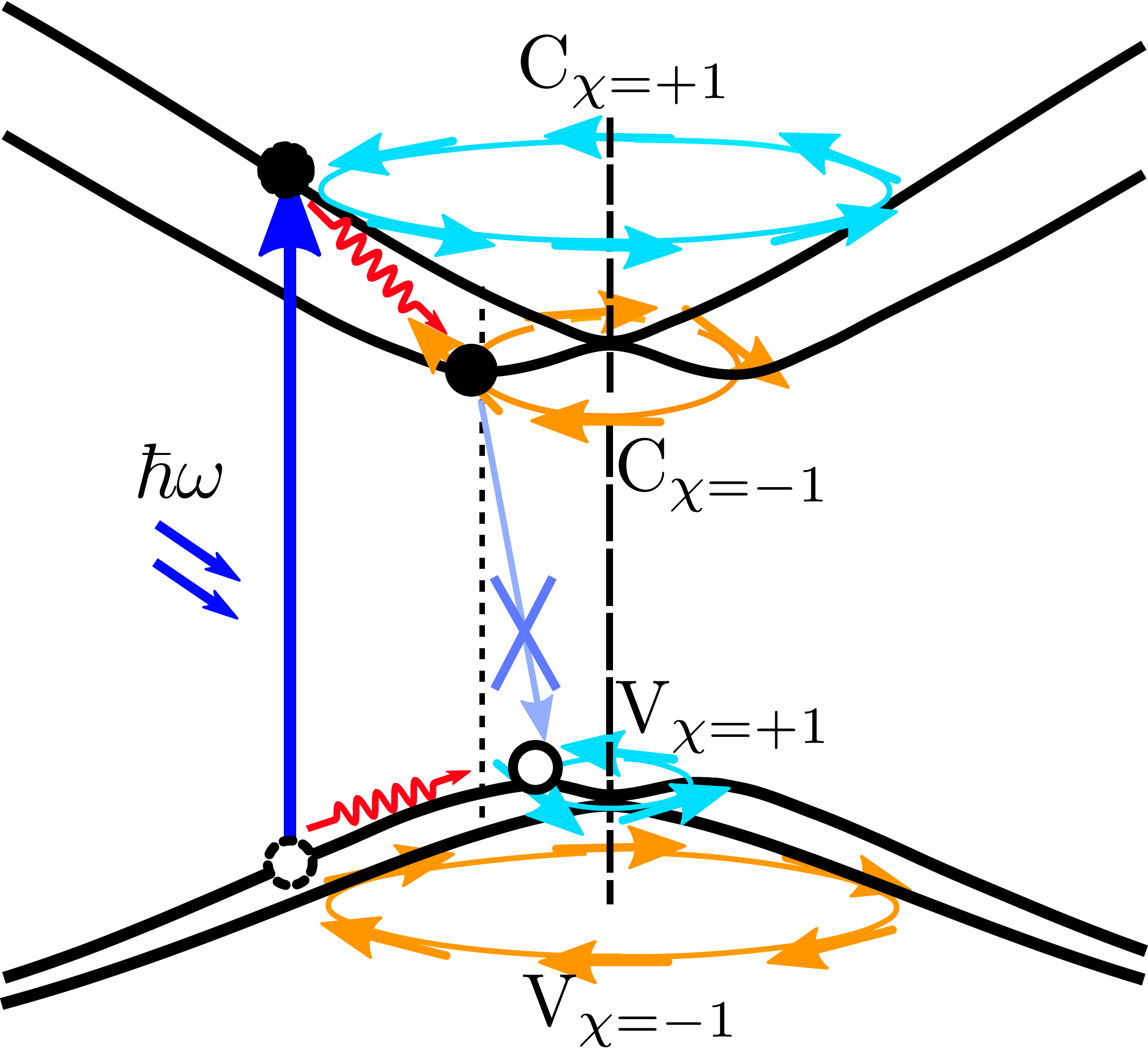}
\caption{\label{band_diagram}Diagram of Rashba bands and the electron transport path. The
cyan and orange arrows indicate the directions of the spins. The spin texture $\chi$ indicates spin vortex direction, with its signs characterizing spin rotation in ``clockwise" ($\chi=-1$) and ``counterclockwise" ($\chi=+1$). After absorbing the photons, the excited electrons on conduction bands ${\rm C}_{\chi=+1}$ and
${\rm C}_{\chi=-1}$ will quickly relax to ${\rm C}_{\chi=-1}$ band minimum due to the
inelastic phonon scattering. Similarly, the holes will quickly relax to the
${\rm V}_{\chi=+1}$ band maximum. However, the radiative
recombination of ${\rm C}_{\chi=-1}$ $\to$ ${\rm V}_{\chi=+1}$ is a spin-forbidden
process due to the opposite spin states they have. Moreover, 
the minimum of ${\rm C}_{\chi=-1}$ band and the maximum of ${\rm V}_{\chi=+1}$ band are located in different positions in the Brillouin Zone. This creates an indirect band gap for
recombination, which further slows down the recombination process.} \end{figure}

Fig.~\ref{band_diagram} illustrates the mechanism for enhancing the carrier lifetime in a generic 3D Rashba material.
The strong spin-orbit coupling effect from heavy elements (e.g., Pb, Sn, I and Br) 
and the polar distortion (e.g., aligned molecular dipoles in OMHPs)
give rise to the Rashba effect, which lifts the two-fold degeneracy
of bands near the band gap.  Near the band gap, the spin degeneracies of the
conduction and valence bands are lifted, giving rise to ``inner" and ``outer" bands with 
opposite spin textures, characterizing spin rotation direction as ``clockwise" ($\chi=-1$) and ``counterclockwise" ($\chi=+1$) (Fig.~\ref{band_diagram}). 
The photo excitation process creates free electrons and holes, which can quickly relax to band extrema in the presence of inelastic phonon scattering.
When the spin textures of conduction band minimum (CBM) 
and valence band maximum (VBM) are opposite, the radiative
recombination of ${\rm C}_{\chi=-1}$ $\to$ ${\rm V}_{\chi=+1}$ is a spin-forbidden
transition due to the mismatch of spin states. This prevents rapid
recombination as the photon-induced spin-flip is a slow process~\cite{Liu14p9310}.
Moreover, the minimum of the ${\rm C}_{\chi=-1}$ band is slightly shifted compared to
the maximum of the ${\rm V}_{\chi=+1}$ band (momentum mismatch). This creates an indirect band gap for
recombination, which further slows down the recombination process due to
the requirement of a phonon with the right momentum~\cite{Motta15p7026}.
In the following, we use the terms favorable and unfavorable relative spin helicity to describe cases when
the VBM and CBM have opposite and aligned spins, respectively.

\begin{figure} \includegraphics[width=0.9\columnwidth]{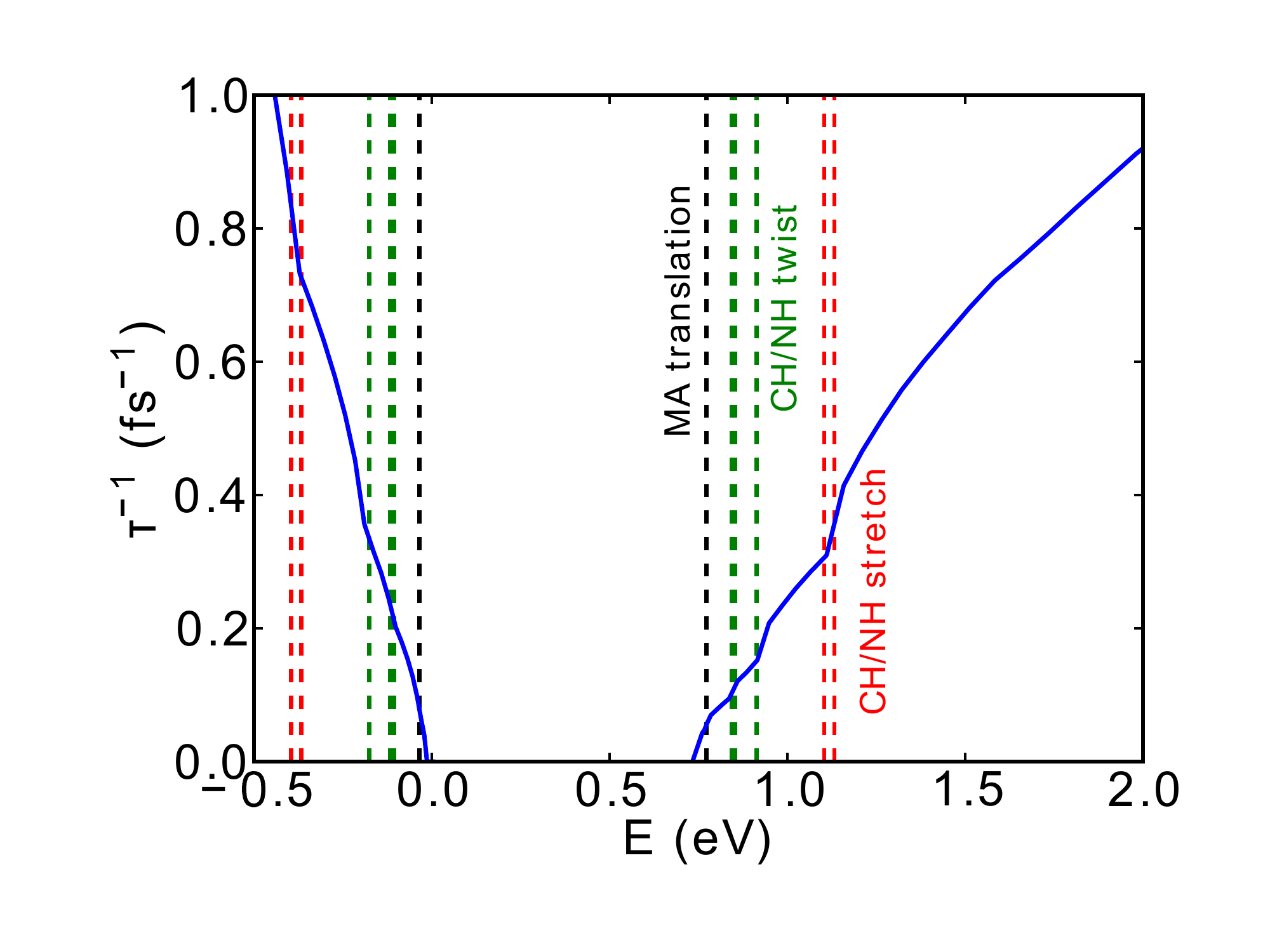}
  \caption{\label{fig_rates} Dependence of phonon-induced relaxation
    rate on carrier energies (blue lines) for electrons (positive
    energies) and holes (negative energies). The VBM is located at $E=0
\textrm{ eV}$, and the CBM is located at $E=0.73 \textrm{ eV}$. The
energies of phonon modes that contribute strongly to carrier relaxation
are shown as dotted lines.  } \end{figure}

The spin texture and carrier population of the CBM and VBM 
play key roles in enhancing carrier 
lifetime. Our first-principles density functional theory (DFT) calculations support 
the realization of this mechanism in OMHPs.
Taking the pseudocubic phase MAPbI$_3$ as an example, we explore the carrier dynamics after
the photoexcitation. 
Using Fermi's golden rule, we calculate the inelastic phonon scattering rate (see Methods).
As shown in Fig.~\ref{fig_rates}, the relaxation rate ($\approx 10^{15}{\rm s}^{-1}$) increases as a
function of carrier energy with sharp jumps corresponding to thresholds
for optical phonon emission. We find that the organic molecule plays an
important role for carrier relaxation. The modes that are most important
for carrier relaxation are MA translation, CH/NH twisting, and CH/NH
stretching, and they contribute equally to the relaxation of both electrons
and holes.\\

Our calculations reveal that the phonon-induced carrier relaxation rate value
is many orders of magnitude faster than the electron-hole recombination
rate ($\approx 10^9$ s$^{-1}$)\cite{Filippetti14p24843,Kawai15pASAP}. 
Therefore, the carriers will rapidly
thermalize and from a quasi-static equilibrium distribution near the CBM
and VBM.  In the ideal case of low temperatures and large Rashba
splitting, nearly all free carriers are located at the band extrema, and
the effects of spin and momentum mismatch on the enhancement of the
carrier lifetime will be the greatest. This effect is less strong at
finite temperatures and small Rashba splitting because of the thermal
occupation of ${\rm C}_{\chi=1}$ and ${\rm V}_{\chi=-1}$ bands, which opens spin-allowed
recombination paths such as  ${\rm C}_{\chi=-1}$ $\to$ ${\rm V}_{\chi=-1}$. In the case of 
favorable relative spin helicity, we investigate
this temperature effect by examining the Rashba splitting using a Rashba
Hamiltonian\cite{Bychkov84p78}:

\begin{equation}\label{rham} 
  H_{R} = \frac{\hbar^2 k^2}{2m} + 
  \hbar v_R\hat{z} \times \vec{k} \cdot \vec{\sigma} 
\end{equation} 
\begin{equation} \label{rdisp}
  E_{k,s} = \frac{\hbar^2 (k_z^2 + k_\bot^2)}{2m}
 +s\hbar v_R k_\bot, \quad k_\bot = \sqrt{k_x^2+k_y^2} 
\end{equation}

\noindent For the parameters $m$ and $v_R$ representing the band mass and Rashba interaction respectively, we use values obtained from the DFT band structure of fully-relaxed MAPbI$_3$.
We find that the conduction band Rashba splitting (0.108
eV) is much larger than the thermal energy scale, while the valence band
Rashba splitting (0.016 eV) is comparable to the thermal energy scale.
Since electronic correlations are not fully captured in DFT~\cite{Filip14p245145}, these values are
likely lower bounds of the true splitting. These relatively large Rashba splittings are likely to give rise to a significant enhancement in carrier lifetime even at room temperature.

\begin{figure} \includegraphics[width=0.5\columnwidth]{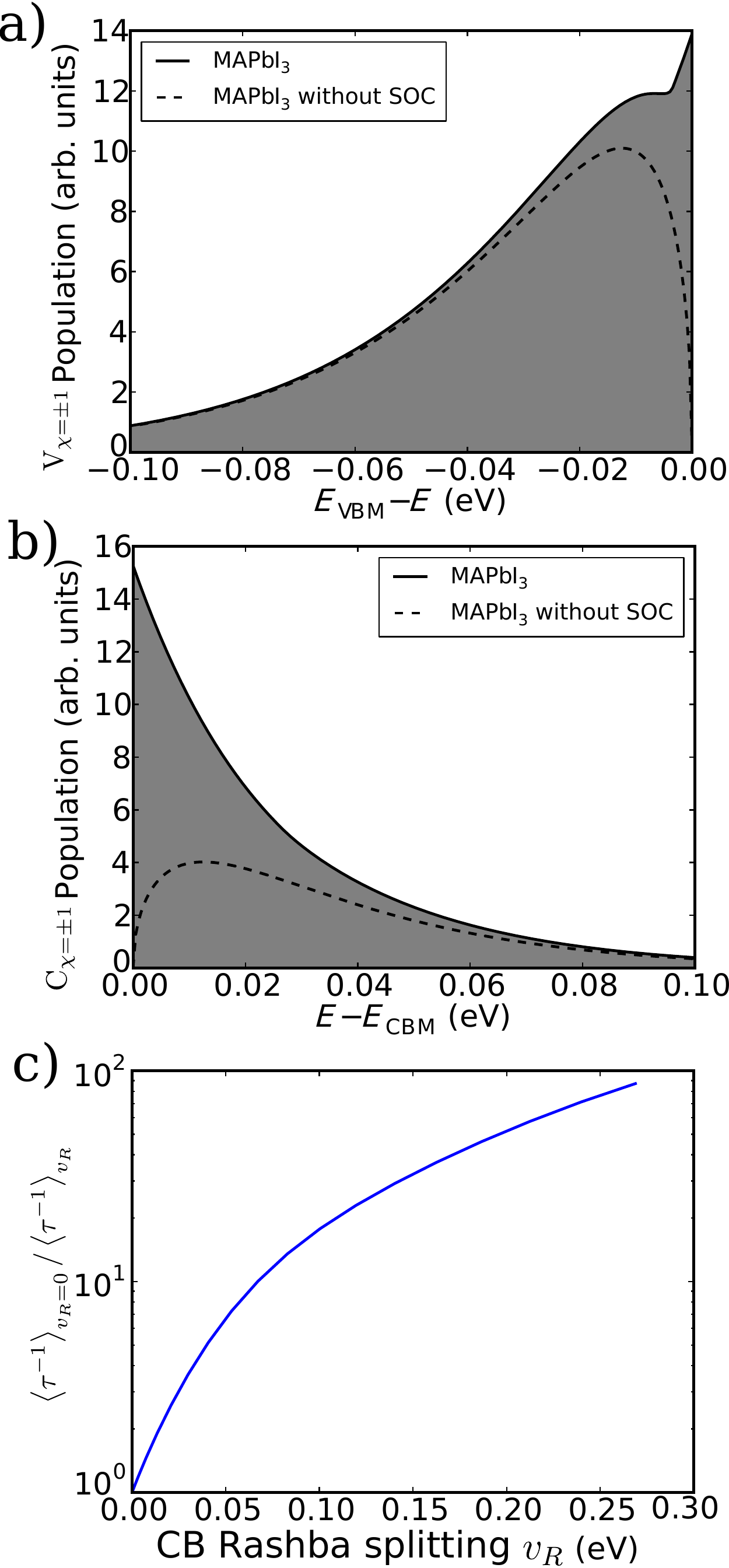}
  \caption{\label{fig_pop} a) Population of carriers at $T$=298 K, calculated
    from the Fermi-Dirac distribution, for a) the top valence band and b) the
bottom conduction band of MAPbI$_3$. Shown in dashed lines are the population of
carriers in a model material with the same band masses as MAPbI$_3$,
but with vanishing Rashba splitting. c) Unitless lifetime enhancement factor,
as defined in the text, as a function of the conduction band Rashba
splitting energy.} \end{figure}

There is an additional factor, arising from the unique features of Rashba band
structure, which promotes occupation of the band extrema. In contrast to
the band extrema of ordinary parabolic bands in semiconductors which are
points in momentum space, those of Rashba materials are one-dimensional
rings (Fig.~\ref{band_diagram}).  This leads to an increase in
the density of states at low energies, resulting in a population of carriers
heavily skewed towards the band extrema (Fig.~\ref{fig_pop}a and b) and
consequently reduces the overall recombination rate due to the reasons of
spin and momentum mismatch as discussed above.

The magnitude of Rashba splitting depends on the amount of polar
distortion and the strength of the SOC, both of which can be
captured by the Rashba velocity parameter $v_R$. We calculate the
averaged recombination rate

\begin{equation} \label{eq3}
  \langle \tau^{-1} \rangle = \sum\limits_{\chi,\chi^\prime} 
  \frac{ \int d^3k \, \tau^{-1}_{\chi,\chi^\prime}(k) \, n^e_{\chi}(k) \, n^h_{\chi^\prime}(k)}
  { \int d^3k \, n^e_{\chi}(k) \, n^h_{\chi^\prime}(k)} 
\end{equation}

\noindent where $\tau^{-1}_{\chi,\chi^\prime}(k) = B_{\chi,\chi^\prime} n^e_{\chi}(k) n^h_{\chi^\prime}(k)$ is
the band- and momentum-resolved recombination rate. The spin-mismatch
effect is captured by the rate constant $B_{\chi,\chi^\prime}$, which is larger
when $\chi$ and $\chi^\prime$ have parallel spins than otherwise.  Enhancement of density of states
enters via the temperature dependent electron and hole 
occupation numbers $n^e(k)$ and $n^h(k)$, which tend to peak at 
different $k$ points because of momentum mismatch.
In order to quantify the effect of Rashba splitting on the recombination
rate, we define the unitless lifetime enhancement factor as the
ratio $\langle \tau^{-1} \rangle_{v_R=0} / \langle \tau^{-1}
\rangle_{v_R}$, where $\langle\tau^{-1}\rangle_{v_R}$ is the average
lifetime when SOC is taken into account, and 
$\langle\tau^{-1}\rangle_{v_R=0}$ refers to 
a calculation where SOC is explicitly set to zero in the Rashba model (Fig.~\ref{fig_pop}c).
Upon tuning the Rashba splitting continously in our model, we find that the lifetime enhancement factor
increases approximately exponentially with Rashba splitting - a consequence of the exponential
behavior of carrier occupation numbers near the tail of the Fermi-Dirac distribution. Our model predicts that a Rashba spliting of 0.1 eV can give rise to an order of magnitude enhancement of carrier lifetime.

\begin{figure} \includegraphics[width=1.0\columnwidth]{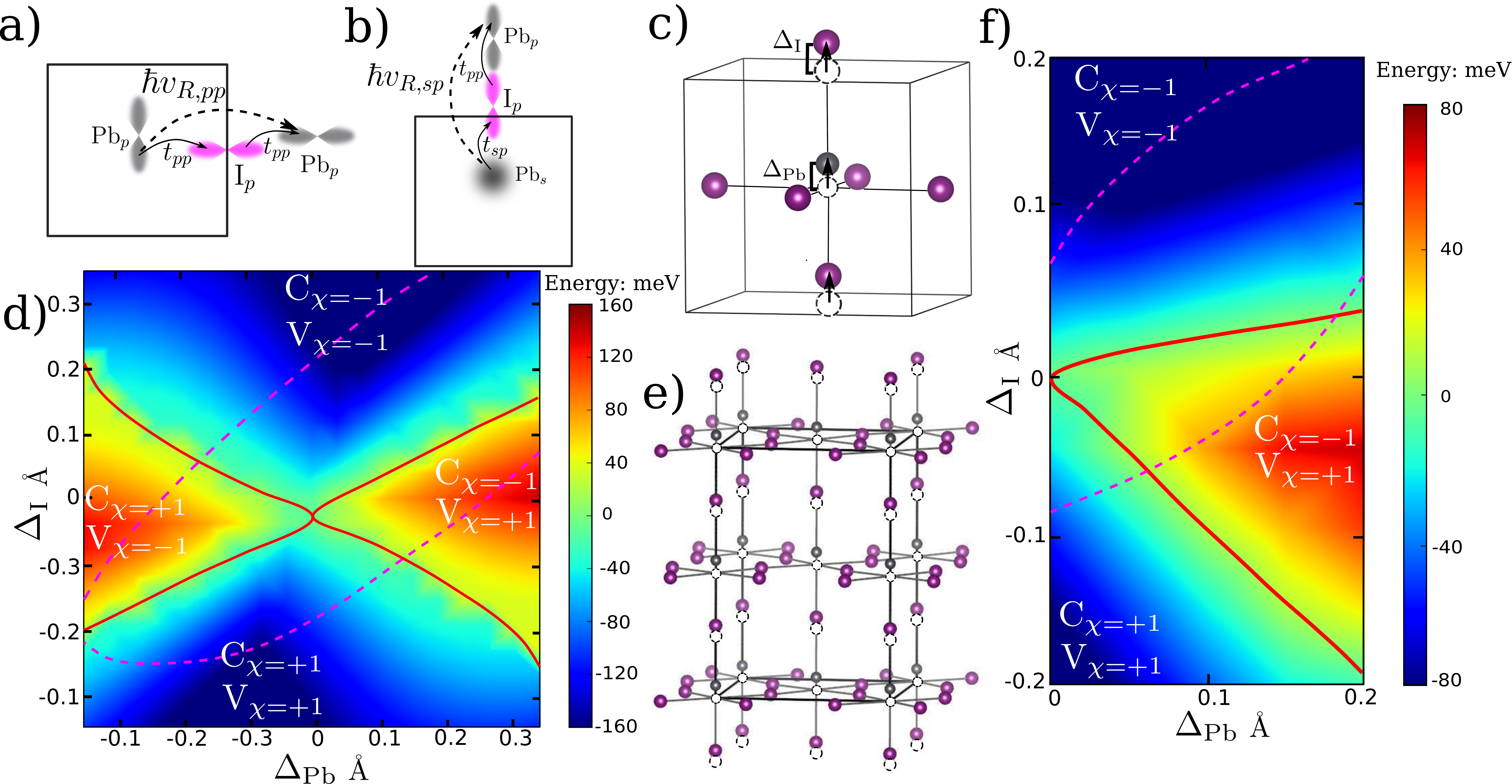}
\caption{
a) and b) Hopping schemes illustrate the effective electric field $\hbar v_{R,sp}$ and $\hbar v_{R,pp}$ created by vertical hopping and horizontal hopping respectively for a range of Pb and I displacements. These two factors caused by different Pb and I displacements controls spin textures of CBM and VBM differently, giving rise to different spin helicities.
c) Schematic diagram showing Pb and I displacement in pseudocubic MAPbI$_3$. Pb: Silver. I: Indigo. Broken circles are original high-symmetry positions. Molecules are not shown here.
d) Phase diagram of splitting energy and spin texture for structures with 
different Pb and apical I displacement in pseudocubic MAPbI$_3$ calculated from DFT. The color is 
the minimum value between the averaged splitting energy of two Rashba conduction bands 
and two valence bands (see Methods). 
The spin texture phase boundaries are indicated by the solid red lines. When the 
structure transforms from a favorable spin texture region to an unfavorable spin texture region, 
the two Rashba valence bands or conduction bands exchange, creating negative 
splitting energy. The dashed lines indicate the areas with energy cost less than 25 meV 
(under room temperature fluctuation) to distort Pb and I. 
e) Schematic diagram showing Pb and I displacement in tetragonal MAPbI$_3$.
f) Similar to d, phase diagram of splitting energy and spin texture for structures with 
different Pb and apical I displacement in tetragonal MAPbI$_3$ calculated from DFT.}\label{2d} \end{figure}

As we have seen, the presence lifetime enhancement is enabled by the relative spin helicity of VBM and CBM, while the amount of enhancement depends on the magnitude of the Rashba splitting. The symmetry-breaking distortions that influence relative spin helicity and splitting magnitude are therefore intimately related to the SOC enhancement of carrier lifetime. To reveal the relation, we start with a tight binding model of the inorganic PbI$_3^{-}$ lattice (see Methods). The displacements of Pb atoms along the $z$ direction give rise to
effective hoppings between Pb $s$- and $p$-orbitals along the equatorial direction (Fig.~\ref{2d}a), which would vanish by symmetry in the absence of such displacements.
Similarly, displacement of the apical I atoms along the $z$ direction changes Pb--I bond lengths
and effective hoppings along the apical direction (Fig.~\ref{2d}b). These modifications of hopping parameters create the effective inversion symmetry breaking electric fields described by $\hbar v_R$ 
in our low-energy theory model (Eq. 1). 
In this model, we find that the spin textures of the valence bands and conduction bands depend on the combination of Pb and I displacements. By shifting Pb and I atoms, we can control the spin textures of the valence and conduction bands, creating favorable and unfavorable relative spin helicities. As we now proceed to show, this picture is confirmed with DFT calculations.

We focus on two phases of MAPbI$_3$, pseudocubic phase 
($\alpha$ phase) and tetragonal phase ($\beta$ phase)~\cite{Stoumpos13p9019}.
The pseudocubic phase of MAPbI$_3$ has space group $P4mm$ with $a=b=6.31\;{\rm \AA}$ 
and $c=6.32\;{\rm \AA}$, which can only hold ferroelectric distortions, such 
as displacements of Pb and apical I atoms. Shown in Fig.~\ref{2d}d are the DFT-calculated 
spin textures and averaged Rashba splittings at VBM and CBM respectively for a given pair of Pb and I displacements (see Methods).
The areas outlined 
by the solid red lines indicate the structures with favorable relative spin helicities, 
which also have relatively large band splittings. We find that these structures have
Pb and apical I displaced along opposite directions. This is consistent with typical
ferroelectric distortions in inorganic ferroelectrics such as BaTiO$_3$ 
and PbTiO$_3$. It is noted that large apical I displacement tends to drive the system away from the region with favorable relative spin helicity to unfavorable spin helicity (Fig.~\ref{2d}d). 

We further explore the relationship between relative spin helicity and local distortions 
in tetragonal MAPbI$_3$~\cite{Stoumpos13p9019}, which is observed at room temperature. The space group 
of the tetragonal phase is identified as $I4cm$, allowing both ferroelectric 
distortion and PbI$_6$ octahedral rotation (Fig.~\ref{2d}e). Fig.~\ref{2d}f shows the spin textures and the averaged Rashba splitting for different Pb and I displacements in $I4cm$ space group. 
The tetragonal phase exhibits similar displacement-helicity relationship to the 
cubic phase, indicating that the Rashba splitting can also enhance the carrier 
lifetime in tetragonal phase.

At room temperature, switchable, large ($\approx$100 nm) 
ferroelectric domains have recently been observed experimentally in MAPbI$_3$~\cite{Kutes14p3335}.
Various theoretical studies also suggested the existence of local polar regions at room 
temperature~\cite{Ma15p248,Frost14p2584,Quarti14p6557}. The structures that are energetically accessible at room temperature are highlighted by dashed red lines in Fig.~\ref{2d}d and f, which cover a large region displaying favorable spin textures. Hence, our proposed mechanism is likely to be realized in MAPbI$_3$ at room temperature, and provides a possible explanation of long carrier lifetime. Moreover, recent studies of carrier dyanmics under magnetic field also illustrate the significance of spin in carrier  recombinations~\cite{Giovanni15p1553,Hsiao15pASAP,Zhang15pASAP}.

Rashba SOC enhanced carrier lifetime highlights the potential of 3D Rashba materials for photovoltaic applications. The ability to incorporate different organic molecules in OMHPs provides a robust avenue to design 3D Rashba materials. 
By tuning the dipole magnitude of organic molecules in halide perovskites, the spin helicities and band splittings can be manipulated via the Pb and I displacements as discussed above.
Conventional experimental techniques of controlling bulk polarization (e.g, epitaxial strain) can also be applied to optimize power conversion efficiency.

In summary, we have proposed an intrinsic mechanism for enhancing carrier
lifetime in 3D Rashba materials. In the case of OMHPs, such mechanism can be realized by the joint action of molecules (electron-phonon coupling) and 
PbI$_3$ sublattices (giving rise to spin-orbit coupling). The photoexcited carriers
quickly relax to band edges due to the electron-phonon coupling. When the 
spin textures for CBM and VBM are opposite, the
Rashba splitting of bands close to the band gap results in spin-allowed
and spin-forbidden recombination paths. The spin-forbidden recombination
path has slow transition rate due to mismatch of spin and momentum. The
spin-allowed recombination path, though kinetically favorable, will only
influence a smaller amount of carriers. In order to achieve this favorable
spin helicity, we explore different Pb and I displacement giving rise 
to different spin textures. A tight binding model is developed to explain 
this spin-displacement relation.
This mechanism allows OMHPs to
behave like direct-gap semiconductors upon photoexcitation, and like
indirect-gap semiconductors during radiative recombination,
simultaneously harnessing
the large carrier densities of the former and the long lifetimes of the
latter. Our work is the first  to introduce this mechanism and highlights the
importance of Rashba effect and structural distortion for achieving long carrier lifetime and
consequently long diffusion length in organometal halide perovskites.

\noindent{\bf {\Large Methods}}\\
\noindent{\bf DFT and Electron-phonon coupling} The plane-wave DFT package QUANTUM-ESPRESSO~\cite{Giannozzi09p395502} with the Perdew-Burke-Ernzerhof~\cite{Perdew96p3865} functional is used to perform electronic structure and electron-phonon coupling calculation. Norm-conserving, designed nonlocal pseudopotentials were
generated with the OPIUM package~\cite{Rappe90p1227,Ramer99p12471}. We calculate the inelastic phonon scattering
rate for electrons and holes using Fermi's golden rule
\cite{Bernardi14p257402,Kawai15pASAP}:
\begin{equation} \tau^{\left(ph\right)\;-1}_{k,n\rightarrow n'} = \hbar^{-1}
\frac{V}{(2\pi)^3} \sum\limits_{\nu} \int d^3k' \lvert
g_{\nu,k,k'}^{n,n'}\rvert^2 \delta(E_{k,n}-E_{k',n'}-\hbar\omega_{\nu})
\end{equation}
\noindent where $k$ and $n$ are the wavevector and band index respectively, $\nu$ denotes the phonon mode, $E_{k,n}$ are electronic band energies and $V$ is the volume of the unit
cell. The phonon frequencies $\omega_\nu$ and electron-phonon
matrix elements $g_{\nu,k,k'}^{n,n'}$ are obtained from density
functional perturbation theory (DFPT)\cite{Baroni01p515}.

\noindent {\bf Lifetime enhancement factor} The rate constant $B_{\chi,\chi^\prime}$ in Eq. 3 of spin-allowed ($\chi=\chi^\prime=\pm1$) and spin-forbidden ($\chi\ne\chi^\prime$) transitions are obtained from averaged DFT calculated oscillation strength. For example, $B_{1,1}\approx 3.7B_{1,-1}$.

\noindent{\bf Tight-binding model} The tight-binding model is based on PbI$_3^{-}$ structure. Pb 6$s$, Pb 6$p$ and the I 5$p$ orbitals are included in this model with spin degree of freedom. The tight-binding Hamiltonian is:
\begin{equation} 
 H_{\rm TB} = H_{\rm hop} + H_{\rm SOC}
\end{equation}
\noindent where $H_{\rm hop}$ considers the nearest neighbor hopping between two orbitals, described by $t_{sp}$, $t_{pp\sigma}$ and $t_{pp\pi}$ for $s$--$p$ $\sigma$ hopping, $p$--$p$ $\sigma$ hopping and $p$--$p$ $\pi$ hopping respectively. $H_{\rm SOC}$ is the on-site SOC term defined as $\lambda_{\rm Rashba}\bv{L\cdot S}$. Hopping parameters and $\lambda_{\rm Rashba}$ are fitted to DFT band structures of pseudocubic MAPbI$_3$ with experimental lattice constants.
We reduce our tight-binding Hamiltonian into the Rashba effective model (Eq. 1) in two steps, following the procedure outlined in~\cite{Kim14p6900}. First, the I $p$ orbitals are removed by projecting $H_{\rm TB}$ to the subspace of Pb orbitals:

\begin{equation}
 H_{\rm Pb} = \mathcal{P} H_{\rm TB} \mathcal{P} + \mathcal{P} H_{\rm TB} \mathcal{Q} \frac{1}{E-\mathcal{Q}H_{\rm TB}\mathcal{Q}} \mathcal{Q} H_{\rm TB} \mathcal{P}
\end{equation}

\noindent
where $\mathcal{P}$ and $\mathcal{Q}$ are projection operators to the Pb and I subspaces respectively. This results in an effective Hamiltonian $H_{\rm Pb}$ containing inversion symmetry breaking terms which modify the effective hopping between Pb orbitals \cite{Kim14p6900}. Next, $H_{\rm Pb}$ is reduced to $H_{R}$ using a similar projection to the CBM and VBM.


\noindent{\bf Rashba splitting energy phase diagram} The Rashba splitting energy is defined as $|\Delta E^{R}| = {\rm min}\left[ \left<E_{{\rm C},s={\rm CBM}+1}-E_{{\rm C},s={\rm CBM}}\right>, \left<E_{{\rm V},s={\rm VBM}}-E_{{\rm V},s={\rm VBM}-1}\right>\right]$. Here, the sign of $\Delta E^R$ is indicated as $+1$ for favorable spin helicity, and $-1$ for unfavorable spin helicity. ``$\left<\;\right>$'' indicates the average over $k$ points near CBM or VBM.

\providecommand{\latin}[1]{#1}
\providecommand*\mcitethebibliography{\thebibliography}
\csname @ifundefined\endcsname{endmcitethebibliography}
  {\let\endmcitethebibliography\endthebibliography}{}


\newpage \end{document}